\documentclass[apj]{emulateapj}
\usepackage{natbib}

\newcommand{\fluxunits}{\mbox{erg cm$^{-2}$ s$^{-1}$}}

\newcommand{\nvw}{\mbox{\ion{N}{5} $\lambda$1240}}
\newcommand{\siivw}{\mbox{\ion{Si}{4} $\lambda$1398}}

\newcommand{\civw}{\mbox{\ion{C}{4} $\lambda$1549}}
\newcommand{\heiiw}{\mbox{\ion{He}{2} $\lambda$1640}}

\newcommand{\oiiw}{\mbox{[\ion{O}{2}] $\lambda$3727}}

\newcommand{\niiw}{\mbox{[\ion{N}{2}] $\lambda \lambda$6548,6583}}

\newcommand{\siiw}{\mbox{[\ion{S}{2}] $\lambda \lambda$6716,6731}}
\newcommand{\oiiiaw}{\mbox{[\ion{O}{3}] $\lambda$4959}}
\newcommand{\oiiibw}{\mbox{[\ion{O}{3}] $\lambda$5007}}
\newcommand{\oiiiw}{\mbox{[\ion{O}{3}] $\lambda \lambda$4959,5007}}

\newcommand{\lya}{\mbox{Ly$\alpha$}}

\newcommand{\heii}{\mbox{\ion{He}{2}}}

\newcommand{\hal}{\mbox{H$\alpha$}}

\newcommand{\hb}{\mbox{H$\beta$}}

\newcommand{\dbrk}{\mbox{$D(4000)$}}
\newcommand{\wrst}{\mbox{$W_{\lambda}^{\mbox{\tiny rest}}$}}
\newcommand{\wrsti}{\mbox{$W_{\lambda,i}^{\mbox{\tiny rest}}$}}
\newcommand{\wobs}{\mbox{$W_{\lambda}^{\mbox{\tiny obs}}$}}

\newcommand{\simgtr}{\, \raisebox{-.2ex}{$\stackrel{>}{\mbox{\tiny $\sim$}}$} \,}
\newcommand{\simlt}{\, \raisebox{-.2ex}{$\stackrel{<}{\mbox{\tiny $\sim$}}$} \,}

\newcommand{\bootes}{Bo\"{o}tes}
\newcommand{\lamfwhm}{\mbox{$\Delta \lambda_{\mbox{\tiny FWHM}}$}}
\newcommand{\lamblaze}{\mbox{$\lambda_{\mbox{\tiny blaze}}$}}

\newcommand{\fshrt}{\mbox{$f_{\tiny \nu}^{\mbox{\tiny short}}$}}
\newcommand{\flng}{\mbox{$f_{\tiny \nu}^{\mbox{\tiny long}}$}}
\newcommand{\ztit}{\mbox{$z \approx 4.5$}}

\begin{document}

%%%%%%%%%%%%%%
% Title Page %
%%%%%%%%%%%%%%

\slugcomment{\it Accepted to ApJ}

\title{Spectroscopic Properties of the \ztit\ \lya--emitters\altaffilmark{1}}

\author{
Steve Dawson\altaffilmark{2},
James E. Rhoads\altaffilmark{3},
Sangeeta Malhotra\altaffilmark{3},
Daniel Stern\altaffilmark{4},
Arjun Dey\altaffilmark{5},
Hyron Spinrad\altaffilmark{2},
Buell T. Jannuzi\altaffilmark{5},
JunXian Wang\altaffilmark{6},
Emily Landes\altaffilmark{2}
}

\altaffiltext{1}{
Based in part on observations made at the W.M. Keck Observatory, which is
operated as a scientific partnership among the California Institute of
Technology, the University of California, and the National Aeronautics and
Space Administration.  The Observatory was made possible by the generous
financial support of the W.M. Keck Foundation.}

\altaffiltext{2}{
Department of Astronomy, University of California at Berkeley, Mail Code
3411, Berkeley, CA 94720 USA; sdawson@astro.berkeley.edu,
spinrad@astro.berkeley.edu, elandes@astro.berkeley.edu}

\altaffiltext{3}{
Space Telescope Science Institute, 3700 San Martin Drive, Baltimore, MD
21218; san@stsci.edu, rhoads@stsci.edu}

\altaffiltext{4}{
Jet Propulsion Laboratory, California Institute of Technology, Mail Stop
169--327, Pasadena, CA 91109 USA; stern@zwolfkinder.jpl.nasa.gov.}

\altaffiltext{5}{
KPNO/NOAO, 950 N. Cherry Ave., P.O. Box 26732, Tucson, AZ 85726 USA;
dey@noao.edu, jannuzi@noao.edu}

\altaffiltext{6}{
Center for Astrophysics, University of Science and Technology of China,
Hefei, Anhui 230026, China}

%%%%%%%%%%%%
% Abstract %
%%%%%%%%%%%%

\begin{abstract}
We present Keck/LRIS optical spectra of 17 \lya--emitting galaxies 
and one Lyman break galaxy at $z \approx 4.5$
discovered in the Large Area Lyman Alpha (LALA) survey.  The survey has
identified a sample of $\sim 350$ candidate \lya--emitting galaxies at $z
\approx 4.5$ in a search volume of $1.5 \times 10^6$ comoving Mpc$^3$. 
We targeted 25 candidates for spectroscopy; hence, the 18
confirmations presented herein suggest a selection
reliability of 72\%.  The large equivalent widths (median $\wrst \approx
80$ \AA) but narrow physical widths ($\Delta v < 500$ km s$^{-1}$) of the
\lya\ emission lines, along with the lack of accompanying high--ionization
state emission lines, suggest that these galaxies are young systems
powered by star formation rather than by AGN activity. Theoretical models
of galaxy formation in the primordial Universe suggest that a small
fraction of \lya--emitting galaxies at $z \approx 4.5$ may still be
nascent, metal--free objects. Indeed, we find with 90\% confidence that 
3 to 5 of the confirmed sources show $\wrst > 240$ \AA, exceeding the
maximum \lya\ equivalent width predicted for normal stellar populations.
Nonetheless, we find no evidence for \heiiw\ emission in either individual
or composite spectra, indicating that though these galaxies are young,
they are not truly primitive, Population III objects.
\end{abstract}

\keywords{galaxies: high--redshift --- galaxies: evolution}

%%%%%%%%
% Body %
%%%%%%%%

\section{Introduction}
\label{introduction}

\lya\ emission has recently begun to fulfill its long--awaited role as a
tracer of young galaxies in the high--redshift universe.  Although early
predictions based on monolithic collapse models \citep{partridge67}
over--estimated characteristic \lya\ line luminosities by factors of $\sim
100$, the basic insight that \lya\ is a good tracer of young stellar
populations in association with the gas from which they formed remains
valid.  Because \lya\ photons are resonantly scattered by neutral
hydrogen, their effective optical depth to dust absorption in the
interstellar medium is greatly enhanced compared to continuum photons of
slightly different wavelength.  This effect was one of the first proposed
to explain non--detections in \lya\ protogalaxy searches \citep{meier81},
and now that \lya\ emission has been detected in high redshift field
galaxies \citep[e.g.][]{hu96, cowie98, dey98, hu98, pascarelle98, hu99,
steidel00, kudritzki00, rhoads00, fynbo01, dawson01, dawson02, lehnert03, bunker03,
kodaira03}, we can turn the effect to our advantage.  The line is produced
by the interaction of ionizing radiation with hydrogen, and is quenched by
dust. Thus, on the simplest interpretation, \lya--selected samples will likely include galaxies
with hot, young stellar populations and little dust, properties which are
expected in primitive systems where little chemical evolution has yet
occurred.

The Large Area Lyman Alpha (LALA) survey \citep{rhoads00} has recently
identified in deep narrow band imaging a large sample of \lya--emitting
galaxies at redshifts $z \approx 4.5$ \citep{malhotra02}, $z \approx 5.7$
\citep{rhoads01,rhoads03}, and $z \approx 6.5$ \citep{rhoads04}. The
rest frame equivalent widths (\wrst)  of the \lya\ emission measured in
the narrow band images generally exceed the maximum expected for a normal
stellar population, with 60\% of the $z \approx 4.5$ sample showing $\wrst
> 240$ \AA\ \citep{malhotra02}.  Such large equivalent widths suggest that
the \lya--emission is produced in one of two scenarios:  (1)  young (age
$< 10^7$ years) galaxies undergoing star formation in primitive
conditions, where metal--free (or low metallicity) gas results in
stellar populations biased toward massive, UV--bright stars; or (2)
\lya--emission powered by AGN activity.

We report on the spectroscopic confirmation of 18 narrow band--selected
galaxies at $z \approx 4.5$. The narrow physical widths (when resolved,
$\Delta v < 500$ km s$^{-1}$) of the observed \lya--emission lines rule
out conventional broad--lined (Type I) AGN as the central engines of the
\lya--emitters.  Moreover, the general lack of accompanying
high--ionization state emission lines (e.g.\ \nvw, \siivw, \civw, \heiiw),
along with the recent non--detection of the $z \approx 4.5$ sources in
deep {\it Chandra} imaging \citep{malhotra03, wang04},
also rules out the comparatively rarer high--redshift
narrow--lined (Type II) AGN.  These findings leave massive star formation in 
low metallicity gas as the likely \lya\ emission mechanism.  The
tantalizing limit of such systems --- star formation in zero--metallicity
gas --- represents the first bout of star formation in the pre--galactic
Universe, and would be recognizable in optical spectroscopy by weak
\heiiw\ emission.

We describe our imaging and spectroscopic observations in
\S~\ref{observations}, and we summarize the results of the spectroscopic
campaign in \S~\ref{results}. In \S~\ref{discussion}, we use our
spectroscopic confirmations to update the statistics of the $z \approx
4.5$ population, and we discuss the implications of \civw\ and \heiiw\
non--detections in composite spectra for the possibility of AGN activity
and/or zero--metallicity star formation among the $z \approx 4.5$ sample.
Throughout this paper we adopt a $\Lambda$--cosmology with
$\Omega_{\mbox{\tiny M}} = 0.3$ and $\Omega_\Lambda = 0.7$, and $H_0 = 70$
km s$^{-1}$ Mpc$^{-1}$.  At $z=4.5$, such a universe is 1.3 Gyr old, the
lookback time is 90.2\% of the total age of the Universe, and an angular
size of 1\farcs0 corresponds to 6.61 kpc.

\section{Observations}
\label{observations}

\subsection{Narrow band and Broad band Imaging}
\label{obs_imaging}

The LALA survey concentrates on two primary fields, ``\bootes" (14:25:57
$+$35:32; J2000.0) and ``Cetus'' (02:05:20 $-$04:55; J2000.0).  Each field
is $36 \times 36$ arcminutes in size, corresponding to a single field of
the 8192 $\times$ 8192 pixel Mosaic CCD camera on the 4m Mayall Telescope
at Kitt Peak National Observatory and on the 4m Blanco
Telescope at Cerro Tololo Inter--American Observatory. The $z \approx 4.5$ search uses
five overlapping narrow band filters each with full width at half maximum
(FWHM) $\approx 80$ \AA. The central wavelengths
are $6559$, $6611$, $6650$, $6692$, and $6730$ \AA, giving a total
redshift coverage of $4.37 < z < 4.57$ and a survey volume of $7.4 \times
10^5$ comoving Mpc$^3$ per field. In roughly 6 hours per filter per field,
we achieve 5$\sigma$ line detections in 2\farcs3 apertures of $\approx 2 \times 10^{-17}$ \fluxunits.

The primary LALA survey fields were chosen to lie within the NOAO Deep
Wide Field Survey \citep[NDWFS;][]{jannuzi99}.  Thus, deep NDWFS broad
band images are available in a custom $B_W$ filter ($\lambda_0 = 4135$
\AA, $\hbox{FWHM}=1278$ \AA) and in the Harris set Kron--Cousins $R$ and $I$, as
well as $J$, $H$, $K$, and $K_s$.
The LALA \bootes\ field benefits from additional deep $V$ and
SDSS $z'$ filter imaging. The imaging data reduction is described in
\citet{rhoads00}, and the $z \approx 4.5$ candidate selection is described
in \citet{malhotra02}.  Briefly, candidates are selected based on a
5$\sigma$ detection in a narrow band filter, the flux density of which
must be twice the $R$--band flux density, and
must exceed the $R$--band flux density at the 4$\sigma$ confidence level.
To guard against foreground interlopers, we set a minimum observed
equivalent width of $\wobs > 80$ \AA, and the candidate must not be
detected in the $B_W$--band.

%-----------------------%
% Spectroscopic Results %
%-----------------------%

\begin{deluxetable*}{llcllcll}
\tablewidth{0pt}
\tablecolumns{8}
\tablecaption{Spectroscopic Properties}
\tablehead{\colhead{}                                   &
\colhead{}                                              &
\colhead{\lya\ Flux\tablenotemark{b}}                   &
\colhead{\wrst\tablenotemark{c}}                        &
\colhead{FWHM\tablenotemark{d}}                         &
\colhead{$\Delta v$\tablenotemark{e}}                   &
\colhead{Continuum ($\mu$Jy)\tablenotemark{f}}          &
\colhead{Continuum ($\mu$Jy)\tablenotemark{f}}          \\
\colhead{Target}                                        &
\colhead{$z$\tablenotemark{a}}                          &
\colhead{($10^{-17}$ erg cm$^{-2}$ s$^{-1}$)}           &
\colhead{(\AA)}                                         &
\colhead{(\AA)}                                         &
\colhead{(km s$^{-1}$)}                                 &
\colhead{Blue side}                                     &
\colhead{Red side}}
\startdata
\sidehead{150$\ell$/mm--grating observations\tablenotemark{g}:} \\
\small J020518.1$-$045616 &\small 4.396 &\small 3.79 $\pm$ 0.34 &\small     80$^{+17}_{-13}$ &\small  18.1 $\pm$ 1.1 &\small  $<$ 1140\tablenotemark{i} &\small  0.028 $\pm$ 0.018 &\small 0.126 $\pm$ 0.022 \\
\small J020521.2$-$045920 &\small 4.451 &\small 2.52 $\pm$ 0.22 &\small     79$^{+24}_{-15}$ &\small  16.8 $\pm$ 1.1 &\small  $<$ 1130\tablenotemark{i} &\small  0.030 $\pm$ 0.016 &\small 0.086 $\pm$ 0.019 \\
\small J020525.7$-$045927 &\small 4.491 &\small 2.15 $\pm$ 0.24 &\small     55$^{+13}_{-10}$ &\small  22.0 $\pm$ 1.1 &\small  $<$ 1120\tablenotemark{i} &\small  0.017 $\pm$ 0.016 &\small 0.106 $\pm$ 0.020 \\
\small J142342.3$+$354607 &\small 4.31: & \nodata & \nodata & \nodata & \nodata &\small  0.032 $\pm$ 0.008 &\small 0.152 $\pm$ 0.009 \\
\small J142350.8$+$354512 &\small 4.376 &\small 3.40 $\pm$ 0.18 &\small     82$^{+94}_{-30}$ &\small  15.2 $\pm$ 1.1 &\small  $<$ 1150\tablenotemark{i} &\small  0.021 $\pm$ 0.011 &\small 0.110 $\pm$ 0.085 \\
 \small J142541.7$+$353351 &\small 4.409 &\small 5.83 $\pm$ 0.25 &\small    335$^{+471}_{-125}$ &\small  26.3 $\pm$ 1.1 &\small $<$ 370 &\small  0.041 $\pm$ 0.010 &\small 0.046 $\pm$ 0.034 \\
\small J142542.0$+$353347 &\small 4.400 &\small 0.66 $\pm$ 0.16 &\small    111$^{+201}_{-47}$ &\small  12.9 $\pm$ 1.1 &\small $<$ 1140\tablenotemark{i} &\small  0.009 $\pm$ 0.009 &\small 0.016 $\pm$ 0.012 \\
\sidehead{400$\ell$/mm--grating observations\tablenotemark{h}:} \\
 \small J020432.3$-$050917 &\small 4.454 &\small 1.96 $\pm$ 0.14 &\small     29$^{+3}_{-2}$ &\small  10.2 $\pm$ 0.3 &\small $<$ 370 &\small  0.069 $\pm$ 0.015 &\small 0.182 $\pm$ 0.015 \\
\small J020439.0$-$051116 &\small 4.446 &\small 8.32 $\pm$ 0.18 &\small    291$^{+80}_{-52}$ &\small  11.3 $\pm$ 0.3 &\small $<$ 440 &\small  0.027 $\pm$ 0.020 &\small 0.077 $\pm$ 0.017 \\
\small J020605.8$-$050441 &\small 4.497 &\small 5.98 $\pm$ 0.29 &\small     58$^{+8}_{-6}$ &\small  11.8 $\pm$ 0.3 &\small $<$ 460 &\small  0.166 $\pm$ 0.039 &\small 0.279 $\pm$ 0.035 \\
\small J020611.7$-$050457 &\small 4.489 &\small 3.01 $\pm$ 0.22 &\small    111$^{+70}_{-31}$ &\small   8.6 $\pm$ 0.3 &\small $<$ 280 &\small  0.015 $\pm$ 0.021 &\small 0.073 $\pm$ 0.029 \\
\small J020611.7$-$050627 &\small 4.460 &\small 1.39 $\pm$ 0.20 &\small     96$^{+96}_{-33}$ &\small   8.3 $\pm$ 0.3 &\small $<$ 260 &\small  0.044 $\pm$ 0.019 &\small 0.039 $\pm$ 0.022 \\
\small J020614.1$-$050032 &\small 4.466 &\small 1.52 $\pm$ 0.20 &\small     22$^{+4}_{-3}$ &\small   7.7 $\pm$ 0.3 &\small $<$ 220 &\small  0.124 $\pm$ 0.028 &\small 0.186 $\pm$ 0.027 \\
\small J142432.6$+$353825 &\small 4.428 &\small 4.33 $\pm$ 0.26 &\small    277$^{+1648}_{-133}$ &\small   7.4 $\pm$ 0.3 &\small  $<$ 190 &\small -0.011 $\pm$ 0.051 &\small 0.042 $\pm$ 0.057 \\
\small J142439.8$+$353801 &\small 4.484 &\small 0.68 $\pm$ 0.42 &\small      $>$ 16\tablenotemark{j}  &\small   7.8 $\pm$ 0.3 & \small $<$ 230 &\small -0.035 $\pm$ 0.059 &\small -0.004 $\pm$ 0.060 \\
\small J142545.5$+$352259 &\small 4.452 &\small 1.42 $\pm$ 0.09 &\small    110$^{+37}_{-23}$ &\small  10.3 $\pm$ 0.3 &\small $<$ 380 &\small  0.010 $\pm$ 0.012 &\small 0.035 $\pm$ 0.009 \\
\small J142624.4$+$353832 &\small 4.457 &\small 2.71 $\pm$ 0.11 &\small     60$^{+6}_{-6}$ &\small   3.3 $\pm$ 0.3 &\small   $<$ 270\tablenotemark{i} &\small  0.078 $\pm$ 0.013 &\small 0.122 $\pm$ 0.013 \\
\small J142628.5$+$353808 &\small 4.407 &\small 6.54 $\pm$ 0.16 &\small    517$^{+584}_{-179}$ &\small   7.7 $\pm$ 0.3 &\small  $<$ 220 &\small -0.094 $\pm$ 0.011 &\small 0.034 $\pm$ 0.021 \\
\enddata
\tablenotetext{a}{The redshift was derived from the wavelength of the peak pixel in the observed line profile.
We estimate the error in this measurement to be $0.002 < \delta_z < 0.004$, depending on the spectroscopic
configuration.  However, we note that this measurement may overestimate the true redshift of the system since
the blue wing of the \lya\ emission is absorbed by foreground neutral hydrogen.}
\label{table_spec_prop}
\tablenotetext{b}{The line flux was determined by totaling the flux of the pixels that fall within
the line profile.  No attempt was made to model the emission line or to account for the very minor contribution of the
continuum to the line.  Quoted
uncertainties account for photon counting errors alone, excluding possible systematic errors.
Despite these caveats, the \lya\ line fluxes measured from the spectra
agree to $2\sigma$ in all but one case with those measured in the narrow band imaging.}
\tablenotetext{c}{The rest frame equivalent widths were determined with $\wrst = (F_{\ell} / f_{\lambda,r}) /
(1+z)$, where $F_{\ell}$ is the flux in the emission line and $f_{\lambda,r}$ is the measured red--side continuum flux
density.  The error bars $\delta w_{\mbox{\tiny +}}$ and $\delta w_{\mbox{\tiny --}}$
are $1 \sigma$ confidence intervals determined by integrating over the probability
density functions $P_i(w)$ described in \S~\ref{ew}.  The error bars are symmetric in probability density--space in the sense that
$\int_{w - \delta w_{\mbox{\tiny --}}}^w P_i(w') \, dw' = \int^{w + \delta w_{\mbox{\tiny +}}}_w P_i(w') \, dw'= 0.34$.}
\tablenotetext{d}{The FWHM was measured directly from the emission line by counting the number of pixels in
the unsmoothed spectrum which exceed a flux equal to half the flux in the peak pixel.  No attempt was made to
account for the minor contribution of the continuum to the height of the peak pixel.}
\tablenotetext{e}{The velocity width $\Delta v$ was determined by subtracting in quadrature the effective
instrumental resolution for a point source, and is therefore an upper limit, as the target may have
angular size comparable to the $\simlt 1\arcsec$ seeing of these data.
Where the emission line is unresolved, the velocity width
is an upper limit set by the effective width of the resolution element itself.}
\tablenotetext{f}{Red and blue side continuum measurements are variance--weighted averages made in 1200 \AA\ wide windows beginning
30 \AA\ from the wavelength of the peak pixel in the emission line.
A small correction factor was subtracted from the variance--weighted averages based on the detection of residual
signal remaining in extractions of source--free, sky--subtracted regions of the two--dimensional
spectra (see text, \S~\ref{obs_spectroscopy}).
Quoted uncertainties account for photon counting errors in the source extractions added in quadrature
to the photon counting errors derived in the blank--sky extractions.}
\tablenotetext{g}{$\lamblaze = 7500$ \AA; 4.8 \AA\, pix$^{-1}$ dispersion; $\lamfwhm \approx 25$ \AA.}
\tablenotetext{h}{$\lamblaze = 8500$ \AA; 1.86 \AA\, pix$^{-1}$ dispersion; $\lamfwhm \approx 6$ \AA.}
\tablenotetext{i}{Line is unresolved.}
\tablenotetext{j}{$2 \sigma$ lower limit.  The measurement of the red--side continuum for this source is formally
consistent with no observable flux.  The equivalent width limit was then set by using a $2 \sigma$ upper
limit to $f_{\lambda,r}$ in the expression given in footnote (c).}
\end{deluxetable*}

\subsection{Spectroscopic Observations}
\label{obs_spectroscopy}

Between 2000 April and 2003 May, we obtained deep spectra of a
cross--section of emission line candidates with the Low Resolution Imaging
Spectrometer \citep[LRIS;][]{oke95} at the Cassegrain foci of the 10m
Keck I and Keck II telescopes (pixel scale 0.215\arcsec pix$^{-1}$).  The observations were divided between two
spectrograph configurations:  low resolution, red channel--only
observations employing the 150 lines mm$^{-1}$ grating ($\lamblaze = 7500$
\AA; 4.8 \AA\, pix$^{-1}$ dispersion; $\lamfwhm \approx 25$ \AA\ $\approx 1000$ km s$^{-1}$), and
higher resolution observations employing simultaneously 
the blue--channel 300 lines mm$^{-1}$ grism ($\lamblaze = 5000$ \AA; 2.64
\AA\, pix$^{-1}$ dispersion; $\lamfwhm \approx$ 14 \AA\ $\approx 600$ km s$^{-1}$) 
and the red--channel 400 lines
mm$^{-1}$ grating ($\lamblaze = 8500$ \AA; 1.86 \AA\, pix$^{-1}$
dispersion; $\lamfwhm \approx 6$ \AA\ $\approx 200$ km s$^{-1}$) 
with a dichroic splitting the channels at 5000 \AA.
The data were taken with slitmasks designed to
obtain spectra for $\sim 15$ targets simultaneously; we employed slit
widths from $1.2\arcsec$ to $1.5$\arcsec.  Total exposure times ranged
from 1.5 hours to 4.3 hours. In each case, the total exposure was broken
into a small number of individual integrations between which we performed
$\sim 3\arcsec$ spatial offsets to facilitate the removal of fringing at
long wavelengths.  The airmass never exceeded 1.3 during the observations,
and the seeing ranged from 0.7\arcsec\ to 1.0\arcsec.
There was no overlap between the objects targeted for 150$\ell$/mm--grating
observations and objects targeted for 400$\ell$/mm--grating observations.

We used the IRAF\footnote{IRAF is distributed by the National Optical
Astronomy Observatory, which is operated by the Association of
Universities for Research in Astronomy, Inc., under cooperative agreement
with the National Science Foundation.} package \citep{tody93} to process
the data following standard slit spectroscopy procedures.  Some aspects of
treating the two--dimensional data were facilitated by a custom
software package, BOGUS\footnote{BOGUS is available online at
http://zwolfkinder.jpl.nasa.gov/$\sim$stern/homepage/bogus.html.}, created
by D. Stern, A.J.  Bunker, and S.A. Stanford.  We extracted
one--dimensional spectra using the optimal extraction algorithm described
in \citet{horne86}. Wavelength calibrations were performed in the standard
fashion using Hg, Ne, Ar, and Kr arc lamps; we employed telluric sky lines
to adjust the wavelength zero--point.  We performed flux calibrations with
longslit observations of standard stars from \citet{massey90} taken with
the instrument in the same configuration as the corresponding slitmask
observation. As the position angle of an observation was set by the desire
to maximize the number of targets on a given slitmask, the observations
were generally not made at the parallactic angle.  Moreover, the data were
generally not collected under photometric conditions: five of our six
observing runs were affected by light to moderate cirrus.

%--------------%
% 400 Profiles %
%--------------%

\begin{figure*}[!t]
\centering
\epsscale{0.9}
\plotone{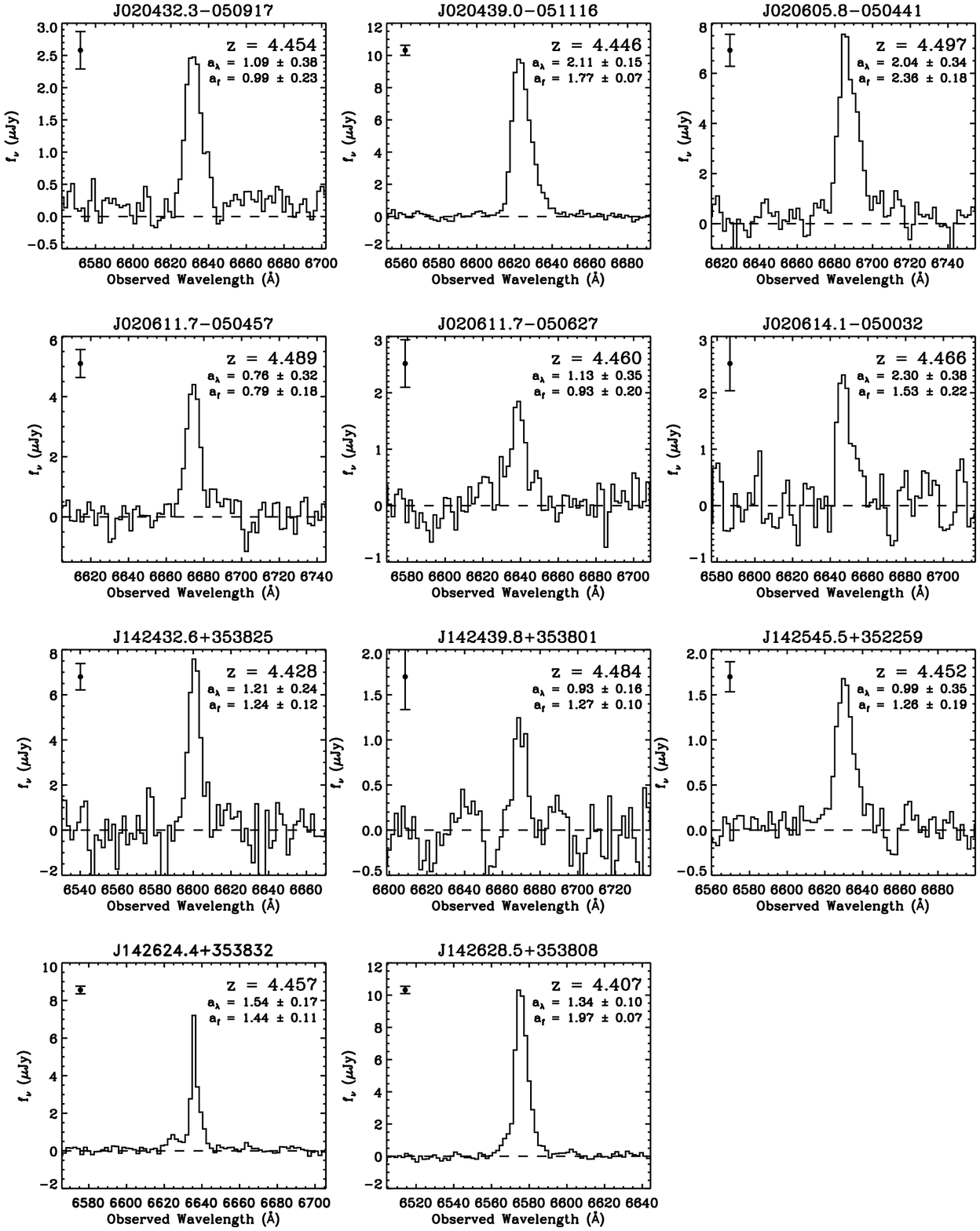}
\caption{Spectra of the 11 confirmed $z \approx 4.5$ galaxies observed with the
Keck/LRIS 400$\ell$/mm--grating, with a wavelength range selected to highlight
the emission line profile.
The measured redshifts and asymmetry statistics (\S~\ref{results}) are indicated at the
upper right.
The representative error bar (upper left) is the median of the flux errors in each pixel
over the wavelength range displayed.  The spectra are unsmoothed.
}
\label{profiles400}
\end{figure*}

To investigate the possibility that the sky--subtraction performed during
object extraction introduced systematic errors, we made twenty extractions
of ostensibly source--free regions in the two--dimensional spectra,
employing the same trace used on the neighboring \lya--emitting candidate.
We then fit the resulting ``blank--sky" spectra in exactly the same way
and over exactly the same wavelength region that we fit for continuum
redward and blueward of the emission line in the object extractions (see
\S~\ref{results} and Table~\ref{table_spec_prop}).  In the complete
absence of systematic errors, we expect the blank--sky fits to be zero,
barring photon counting statistics.  In fact, we find a tiny residual
signal: $f_{\nu} = 0.010$ $\mu$Jy $\pm$ 0.002 $\mu$Jy blueward
of the emission lines, and $f_{\nu} = 0.002$ $\mu$Jy $\pm$ 0.001 $\mu$Jy
redward of the emission lines. These values and their error
bars constitute the weighted--averages of the residual signals measured in
the 20 extractions, where the weights are the uncertainty in each fit.
The error bars represent the uncertainty in the mean residual level
(i.e.\ rather than the scatter among the twenty measurements).
The residual signals were subtracted from the object continuum fits, and the errors were
adjusted accordingly.

\section{Spectroscopic Results}
\label{results}

\begin{figure*}[!t]
\centering
\epsscale{0.9}
\plotone{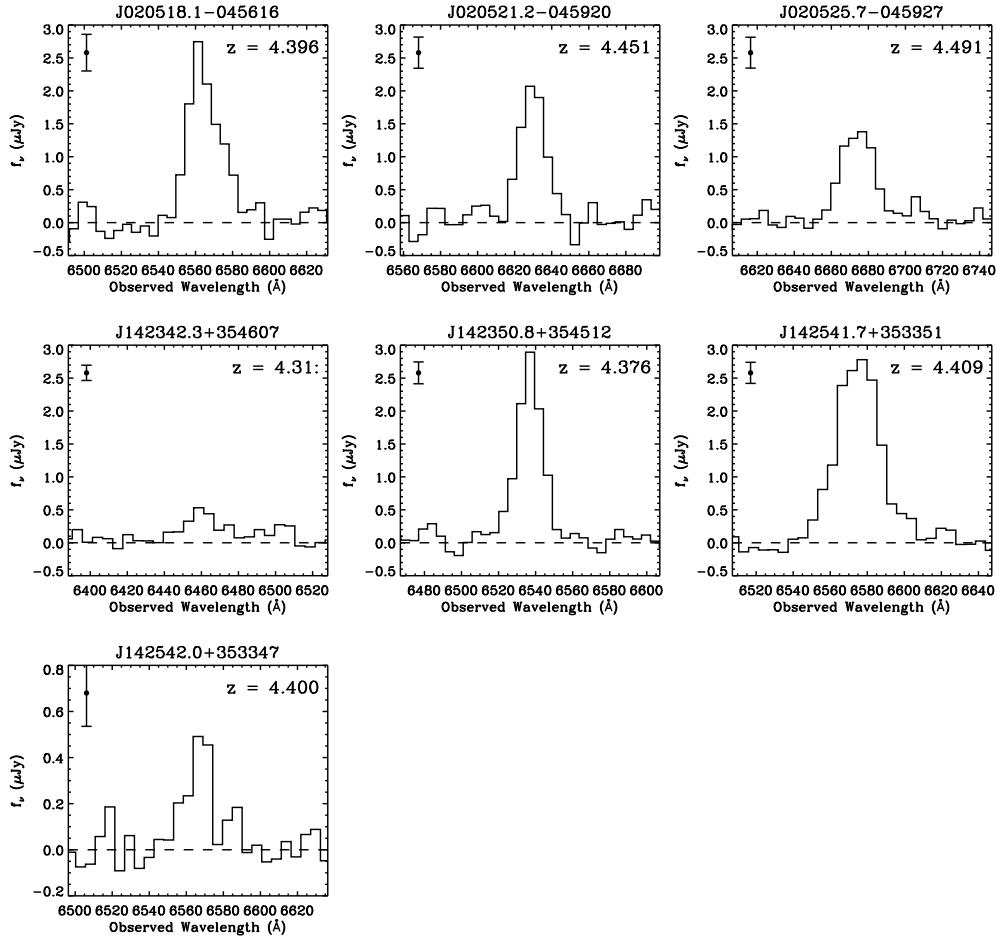}
\caption{Spectra of the 7 confirmed $z \approx 4.5$ galaxies observed with the
Keck/LRIS 150$\ell$/mm--grating, with a wavelength range selected to highlight
the emission line profile.
The measured redshifts are indicated at the upper right.
The representative error bar (upper left) is the median of the flux errors in each pixel
over the wavelength range displayed.
The spectra are unsmoothed.
}
\label{profiles150}
\end{figure*}

Of 25 spectroscopic candidates, 18 were confirmed as galaxies at $z
\approx 4.5$.  All but one of the 18
confirmed galaxies show \lya\ in emission; the remaining galaxy lacks an
emission line but shows a large spectral discontinuity identified as the
onset of foreground \lya--forest absorption at $\lambda_{\mbox{\tiny
rest}} = 1216$ \AA. Of the 7 targets that were not confirmed as
\lya--emitters, 6 were non--detections (to a 5$\sigma$ upper limit of
$\sim 1.1 \times 10^{-17}$ ergs cm$^{-2}$ s$^{-1}$), and one was a clear
\oiiw--emitter at $z=0.801$, based on the presence of continuum blueward
of the emission line. Table~\ref{table_spec_prop} gives the redshifts of
the confirmed \lya--emitters and summarizes the properties of the emission
lines.  Figures~\ref{profiles400} and \ref{profiles150} contain the
confirming spectra.

%--------------%
% 150 Profiles %
%--------------%

We note that each confirmation of a $z \approx 4.5$ \lya--emitter
originates in the spectroscopic detection of a single emission line, the
interpretation of which can be problematic.  In general, a single, isolated
line could be any one of \lya, \oiiw, \oiiibw, or \hal, though given
sufficient spectral coverage most erroneous interpretations can be ruled
out by the presence of neighboring emission lines: \hb\ and \oiiiaw\ for
\oiiibw; \niiw\ and \siiw\ for \hal. Hence, the primary threat to
determining one--line redshifts is the potential for misidentifying \lya\
as \oiiw, or vice versa.  Even so, \oiiw\ at $z = 0.8$ is generally
accompanied by the \hb\ plus \oiiiw\ complex, redshifted to $\approx 9000$
\AA.  Such a detection is challenging, owing to heavy contamination by
night--sky emission lines in this region of the spectrum.  
Nonetheless, each spectrum described herein is consistent with an \oiiibw\
non--detection, with a typical 5$\sigma$ upper limit to \oiiibw\ flux of $\sim 3 \times
10^{-17}$ ergs cm$^{-2}$ s$^{-1}$.  One source has a 5$\sigma$ upper limit of 
$9 \times 10^{-17}$ ergs cm$^{-2}$ s$^{-1}$; the search region for 
\oiiibw\ in this case happens to fall directly on prominent sky line residuals.
Ignoring this outlier, the scatter in the 5$\sigma$ upper limits for
the remaining 17 sources is just $0.5 \times 10^{-17}$ ergs cm$^{-2}$ s$^{-1}$.

%%% Adjusted to satisfy referee
% Nonetheless,
% each spectrum described herein is consistent with an \oiiibw\
% non--detection, with 5$\sigma$ upper limits ranging from $\sim 0.8 \times
% 10^{-17}$ ergs cm$^{-2}$ s$^{-1}$ to $9 \times 10^{-17}$ ergs cm$^{-2}$
% s$^{-1}$, and with 16 out of 18 spectra showing 5$\sigma$ upper limits $<
% 2.3 \times 10^{-17}$ ergs cm$^{-2}$ s$^{-1}$.
%%%

\subsection{Results from the 400$\ell$/mm--Grating Observations}

The confirmed $z \approx 4.5$ sources observed in our higher--resolution
(400$\ell$/mm) spectroscopic configuration typically show the asymmetric
emission line profile characteristic of high--redshift \lya, where neutral
hydrogen outflowing from an actively star--forming galaxy imposes a sharp
blue cutoff and broad red wing \citep[e.g.][]{stern99, manning00,
dawson02, hu04, rhoads03}. The opposite profile is expected for \oiiw\
observed at this resolution \citep[e.g.\ see][]{rhoads03}; hence, the
asymmetry typically detected in our higher--resolution sample is good
evidence for the \lya--interpretation.

%-----------%
% Asymmetry %
%-----------%

\begin{figure*}[!t]
\centering
\epsscale{0.9}
\plotone{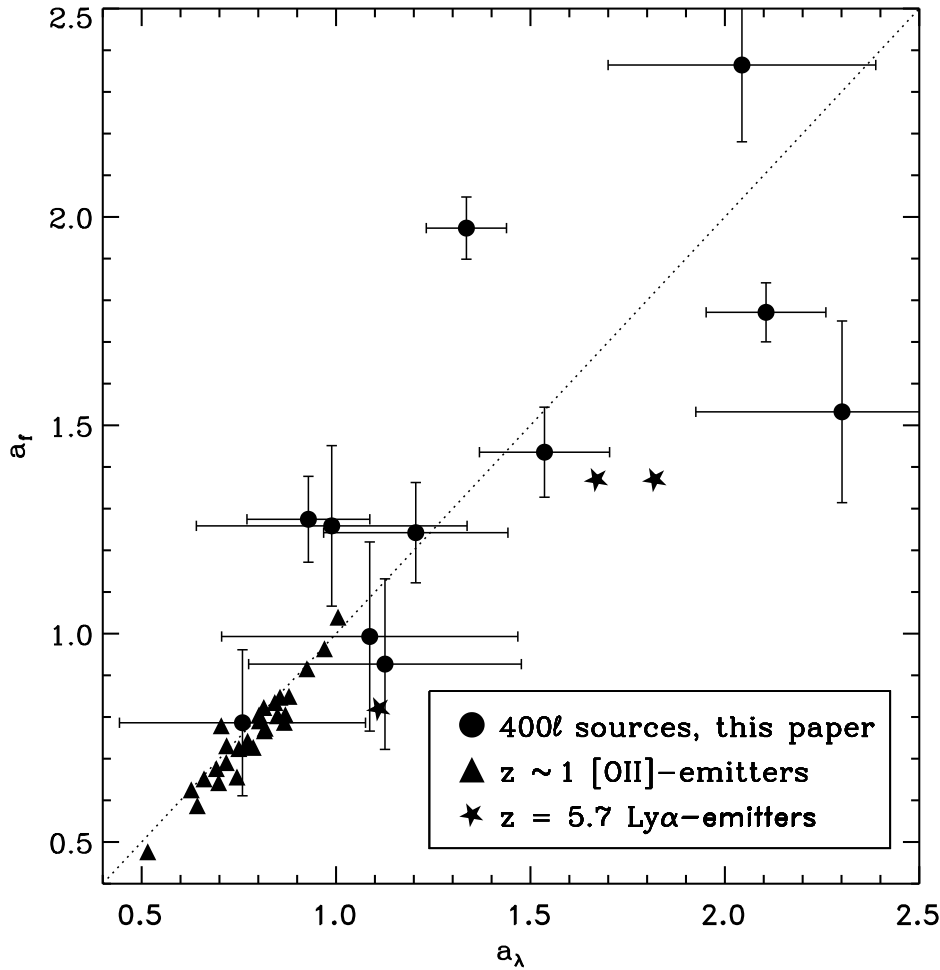}
\caption{Scatter plot of the flux--based asymmetry statistic $a_f$ vs.\ the wavelength--based
asymmetry statistic $a_{\lambda}$ for known high--redshift \lya--emitters, and for a sample
of \oiiw--emitters at $z \sim 1$.
The 11 \lya--emitters at $z \sim 4.5$ are the 400${\ell}$/mm--grating
observations presented in this paper.  (The 150${\ell}$/mm--grating observations,
being of much lower resolution, are not included.)
The 3 \lya--emitters at $z = 5.7$ are from \citet{rhoads03}.
The 28 \oiiw--emitters at $z \sim 1$
were provided by the DEEP2 team \cite[][A.\ Coil 2004, private communication]{davis03}; their Keck/DEIMOS
1200$\ell$/mm--grating spectra were smoothed to the Keck/LRIS 400$\ell$/mm--grating resolution by
convolution with a Gaussian kernel.
}
\label{asym_fig}
\end{figure*}

To quantify this conclusion, we consider two measures of line asymmetry
\citep[see][]{rhoads03,rhoads04}.  For both, we determine the wavelength of the peak
of the emission line ($\lambda_p$), and where the line flux exceeds 10\%
of the peak on the blue side ($\lambda_{\mbox{\tiny 10},b}$) and on the
red side ($\lambda_{\mbox{\tiny 10},r}$). The ``wavelength ratio" is then
$a_{\lambda} = (\lambda_{\mbox{\tiny 10},r} -
\lambda_p)/(\lambda_p - \lambda_{\mbox{\tiny 10},b})$, and the ``flux
ratio" is $a_f = \int_{\lambda_p}^{\lambda_{\mbox{\tiny 10},r}}
f_{\lambda} d \lambda / \int_{\lambda_p}^{\lambda_{\mbox{\tiny 10},b}}
f_{\lambda} d \lambda$\footnote{The error bars on $a_{\lambda}$
and $a_f$ were determined with Monte Carlo simulations in which we
modeled each emission line with the truncated Gaussian profile
described in \citet{hu04} and \citet{rhoads04}, added random noise in each pixel 
according to the photon counting errors, and measured the widths $\sigma(a_{\lambda})$
and $\sigma(a_f)$
of the resulting distributions of $a_{\lambda}$ and $a_f$ for the
given line.  That is, for 
each $a_{\lambda,i}$, the error $\delta a_{\lambda,i} = \sigma(a_{\lambda,i})$,
and similarly for each $a_{f,i}$.}.
As in \citet{rhoads04}, we experimented with raising and lowering the 
flux threshold but found no clear benefit to using values other than 10\%.
Lowering the threshold results in enhanced contamination from continuum noise,
increasing the scatter and uncertainty in the measurements; raising the threshold
diminishes the contribution of the broad, red \lya\ wing to the measurement,
reducing the ability of the asymmetry measures to discriminate \lya\ from
symmetric interlopers.
Figure~\ref{asym_fig} compares
the distribution in $a_f$--$a_{\lambda}$ space of 28 $z \sim 1$
\oiiw--emitters \citep[provided by the DEEP2 team;][A.\ Coil 2004, private communication]{davis03}
to 22 confirmed and putative high--redshift
\lya--emitters.  As a population, the \lya--emitters are systematically
segregated from the \oiiw--emitters. While the \oiiw--emitters are
distributed according to $a_f = 0.8 \pm 0.1$ and $a_{\lambda} = 0.8 \pm
0.1$, all save one of the \lya--emitters satisfies $a_f > 1.0$ or
$a_{\lambda} > 1.0$, and more than half satisfy both.

%-------------------%
% Composite Spectra %
%-------------------%

\begin{figure*}[!t]
\centering
\epsscale{1.0}
\plotone{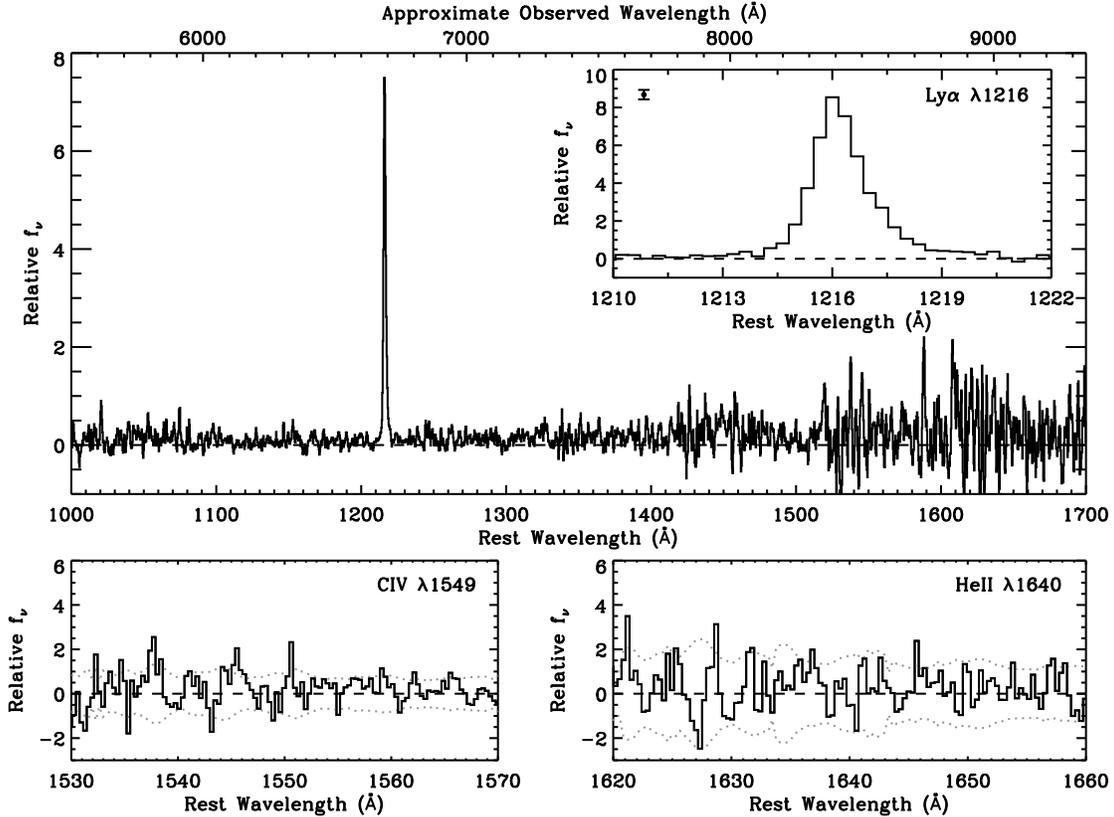}
\caption{
Composite spectrum consisting of an unweighted coaddition of the 11 $z
\approx 4.5$ galaxies confirmed in 400$\ell$/mm--grating observations.  The
full spectrum (top) has been smoothed with a 3 pixel boxcar filter; the
inset, highlighting the asymmetry in the composite \lya\ profile, is
unsmoothed. The representative error bar (upper left in inset) is the
median of the flux errors in each pixel over the wavelength range
displayed.  The small plots at bottom demonstrate the absence of any
significant emission from \ion{C}{4} $\lambda$1549 or \ion{He}{2}
$\lambda$1640 in the composite.
To wit, the \ion{C}{4} $\lambda$1549 flux is
constrained to be $< 8\%$ (12\%) of the flux in \lya\
to a confidence of 2$\sigma$ (3$\sigma$); the \ion{He}{2}
$\lambda$1640 flux is constrained to be $< 13\%$ (20\%) of the flux
in \lya\ to a confidence of 2$\sigma$ (3$\sigma$).
The dotted lines indicate the photon counting errors as they
were propagated through the coaddition.
}
\label{compy400}
\end{figure*}

\begin{figure*}[!t]
\centering
\epsscale{0.9}
\plotone{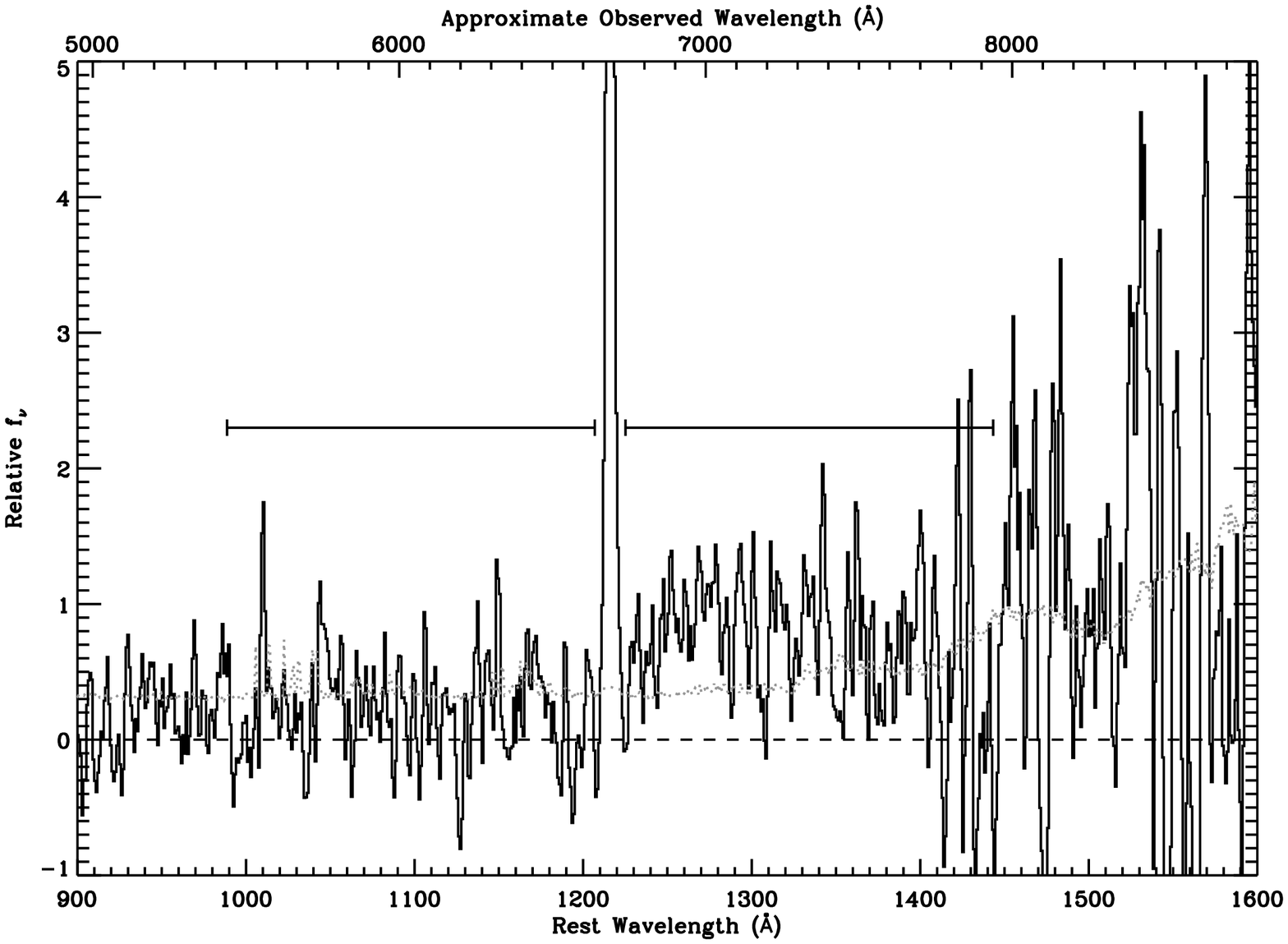}
\caption{Composite spectrum consisting of an unweighted coaddition of the
7 $z \approx 4.5$ galaxies confirmed in 150$\ell$/mm--grating observations,
with the ordinate scale selected to emphasize the continuum break
across the emission line.  The spectrum has been smoothed
with a 3 pixel boxcar filter.  The dotted line indicates
the photon counting errors as they were propagated through the coaddition,
and accounts for the smoothing.  The horizontal bars demarcate the
wavelength region considered for the determination of $1 - \fshrt / \flng$,
which is equal to $0.75 \pm 0.06$ ($1\sigma$) for the composite.
}
\label{compy150}
\end{figure*}

We note that one source included herein as a confirmed \lya--emitter
(J020611.7$-$050457) falls in the region of $a_f$--$a_{\lambda}$ space
characteristic of the \oiiw--emitters. With $a_f = 0.79 \pm 0.10$
($1\sigma$) and $a_{\lambda} = 0.76 \pm 0.05$ ($1\sigma$), the asymmetry
measures do not favor the \lya--interpretation for this source, while the
lack of blue continuum emission, on the other hand, does not favor the
\oiiw--interpretation.  The large equivalent width measured for the line
is similarly ambiguous. For \lya, the redshift is $z=4.489$ and the {\it
observed} frame equivalent width of $\wobs = 620$ \AA\ $\pm$ 200 \AA\
($1\sigma$) implies a rest frame equivalent width of $\wrst = 111$ \AA\
$\pm$ 37 \AA\ ($1\sigma$).  This value is consistent with \wrst\ for other
high--redshift \lya--lines presented here and elsewhere
\citep{hu96,cowie98,hu98,malhotra02,rhoads03}. Alternatively, interpreting
the emission line as \oiiw\ implies a rest frame equivalent width of
$\wrst = 340$ \AA. Such a large value for \oiiw\ is rare but not wholly
unprecedented: large continuum--selected \citep{cowie96,hogg98} and
\hal--selected \citep{gallego96} samples indicate that \oiiw\ very rarely
exceeds $\wrst > 100$ \AA, but at least one exceptional source with $\wrst
\approx 600$ \AA\ at $z=1.464$ is known \citep{stern00}.  
The \oiiw--interpretation would further imply the presence of 
\oiiibw\ at a redshifted wavelength of 8980 \AA; no such line
is detected to a 5$\sigma$ upper limit of $<
2.3 \times 10^{-17}$ ergs cm$^{-2}$ s$^{-1}$.
Though we tentatively include this source among our confirmed \lya--emitters,
the reader is encouraged to keep the foregoing caveat
concerning its interpretation in mind.

\subsection{Results from the 150$\ell$/mm--Grating Observations}

Our lower resolution (150$\ell$/mm) spectroscopic configuration trades
sensitivity to line shape for sensitivity to continuum.  Therefore, while
the (typically unresolved) line profiles in this sample are less useful
for discriminating \lya\ from \oiiw, the continuum detection (if any) can
be used to measure the amplitude of the discontinuity at the emission
line. On the \lya--interpretation, we expect a continuum break owing to
the onset of absorption by the \lya--forest at $\lambda_{\mbox{\tiny
rest}} = 1216$ \AA, a robust spectral signature used extensively in the
photometric selection of galaxies at $z>5$ \citep[e.g.][]{dey98, weymann98,
spinrad98, lehnert03}. The likely low--redshift interloper in this case is
the 4000 \AA\ break [\dbrk], resulting from the sudden onset of stellar
photospheric opacity shortward of 4000 \AA.  We characterize the break
amplitude in the 150$\ell$/mm--grating observations with $1 - \fshrt /
\flng$, where \fshrt\ is the variance--weighted average flux density in a 1200
\AA\ window beginning 30 \AA\ below the emission line, and \flng\ is the
same, but above the emission line.  The median of the 1$\sigma$ {\it
lower} limits to the amplitude of the flux decrement in our $z \approx
4.5$ candidates, calculated from Monte Carlo simulations of the flux
densities with the constraint $f_{\nu} > 0$, is then $1 - \fshrt / \flng >
0.61$.  This value is entirely consistent with both theoretical models of
the \lya\ break amplitude at $z=4.5$ \citep[e.g.][]{madau95, zhang97} and
with measurements from existing data sets \citep[see][and references
therein]{stern99}. By contrast, a sample of 43 galaxies in the redshift
range $0.7 < z < 0.94$ (roughly corresponding to the resulting redshift if
\oiiw\ has been misidentified as \lya) has \dbrk\ amplitudes of $1 -
\fshrt / \flng = 0.39 \pm 0.1$ \citep{dressler90}. We take \lya\ as the
most likely line identification under these circumstances.

We present an unweighted coaddition of the 11 400$\ell$/mm spectra in
Figure~\ref{compy400} and of the 7 150$\ell$/mm spectra in
Figure~\ref{compy150}. With its higher resolution, the 400$\ell$/mm composite
spectrum highlights the asymmetry of the \lya\ emission line.  We find
$a_f = 1.58 \pm 0.09$ ($1\sigma$) and $a_{\lambda} = 1.69 \pm 0.21$
($1\sigma$) for the composite.  With its greater sensitivity to continuum,
the 150$\ell$/mm composite highlights the spectral discontinuity at the
emission line.  We find $1 - \fshrt / \flng = 0.75 \pm 0.06$ ($1\sigma$)
for the composite.  The rest frame equivalent widths of the composite
emission lines are $\wrst(400\ell\mbox{/mm}) = 100$ \AA\ $\pm$ 13 \AA\ ($1\sigma$)
and $\wrst(150\ell\mbox{/mm}) = 74$ \AA\ $\pm$ 13 \AA\ ($1\sigma$), respectively.

\subsection{Spectroscopic Non--Detections}

Of the six spectroscopic non--detections, three candidates fell on
slitmasks observed under adverse conditions for which the general
spectroscopic yield was low.  As such, our failure to confirm these
targets as $z \approx 4.5$ \lya--emitters should not be interpreted as a
reflection on the efficacy of candidate selection. The remaining three
non--detections were observed under acceptable or photometric conditions,
for which the general spectroscopic yield was high.  However, each of
these targets was suboptimal for one of a variety of reasons: one target
sits on a weak satellite residual; one target is very close to a bright
star; one target appears in an initial epoch of imaging but not in 
subsequent epochs and is therefore likely a variable source or a spurious
detection. For the remainder of this paper, we will cite a selection
reliability of 72\%, but the reader is cautioned that this is the most
conservative estimation; it does not discriminate between spurious   
candidates produced at the stage of narrow band--selection, or failures in
the spectroscopic follow--up.  That is, this figure for the selection
reliability does not imply that 28\% of candidate \lya--emitters in narrow
band imaging surveys are unsound.

\section{Discussion}
\label{discussion}

\subsection{The Statistics of the $z=4.5$ Population}
\label{stats}

The spectroscopic confirmation of 17 $z \approx 4.5$ \lya--emitters out of
25 candidates allows us to update the statistics of this population as
they were presented in \citet{malhotra02}.  By applying a $0.72$
correction factor to the observed source counts, we find a number density
of $\approx 2800$ \lya--emitters per square degree per unit redshift above
a detection threshold of $2 \times 10^{-17}$ ergs cm$^{2}$ s$^{-1}$. This
figure translates to a \lya--luminosity density at $z \approx 4.5$ of $2
\times 10^5 L_{\sun}$ Mpc$^{-3}$ for sources with $L_{\mbox{\tiny
Ly$\alpha$}} > 1.04 \times 10^9 L_{\sun}$. \citet{hu99} give a conversion
factor of 1 $M_{\sun}$ yr$^{-1} = 10^{42}$ ergs s$^{-1}$ for converting
\lya--luminosities into star formation rates \citep[but see the caveats
in][]{rhoads03}; together with the \lya--luminosity density estimate, this
yields a star formation rate density at $z \approx 4.5$ of $8 \times
10^{-4} M_{\sun}$ yr$^{-1}$ Mpc$^{-3}$, with individual star formation
rates ranging from 1 $M_{\sun}$ yr$^{-1}$ to 16 $M_{\sun}$ yr$^{-1}$.
These results are roughly consistent with previous studies with similar
limiting fluxes \citep[e.g.][]{hu98,ouchi03}.

%-----------------%
% EW Distribution %
%-----------------%

\begin{figure*}
\centering
\epsscale{1.0}
\plotone{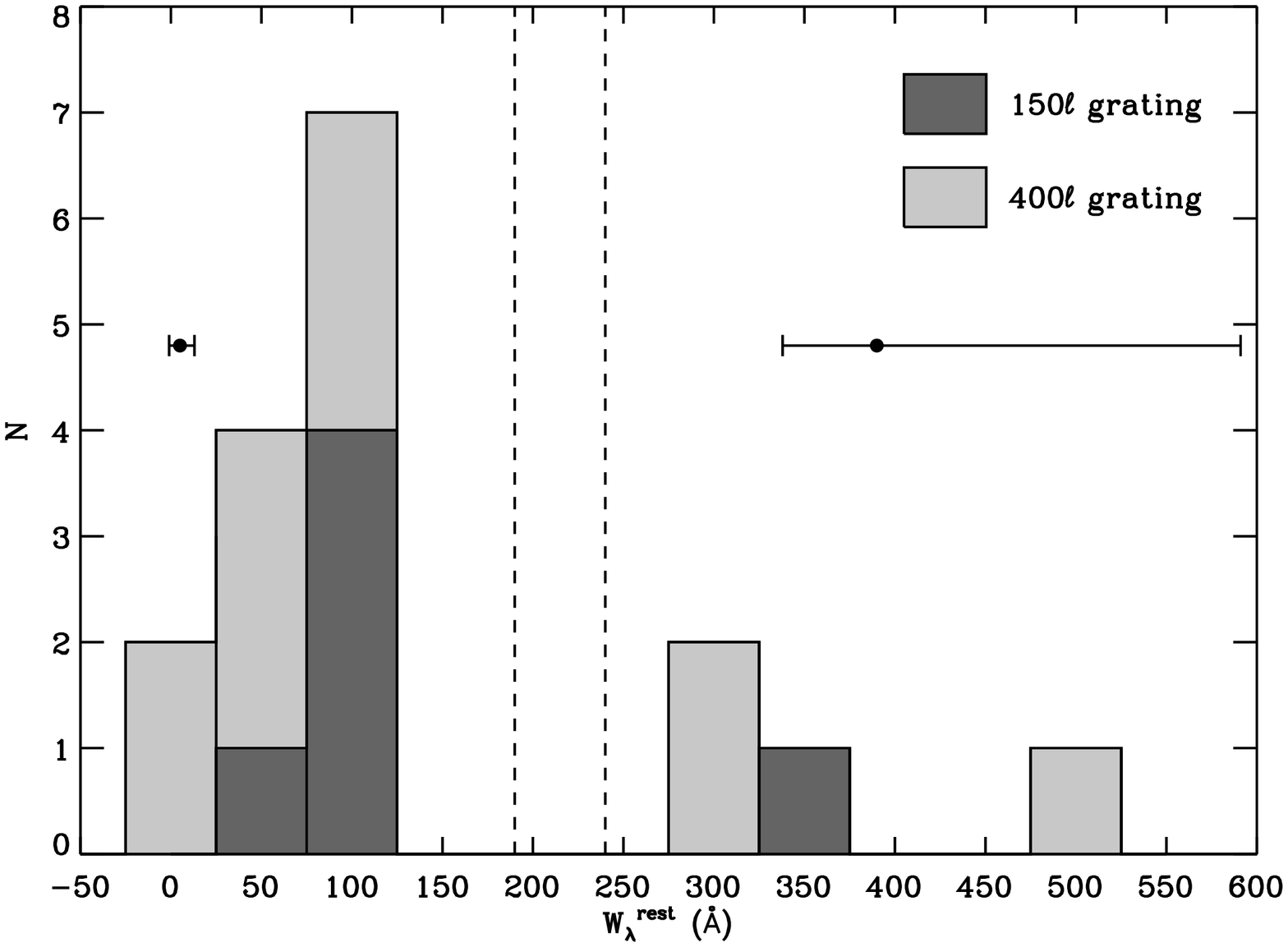}
\caption{
Histogram of the spectroscopic rest frame equivalent widths for the $z=4.5$ population,
determined with $\wrst = (F_{\ell} / f_{\lambda,r}) /
(1+z)$, where $F_{\ell}$ is the flux in the emission line and $f_{\lambda,r}$ is the measured
red--side continuum flux
density.  The dashed vertical lines mark the maximum \lya\ equivalent widths
predicted by the stellar models of \citet{charlot93} and \citet{malhotra02}.
Representative error bars on the equivalent widths are plotted at left and at right.
Notably, the highest equivalent widths are generally the least certain, as they
correspond to the faintest (and hence least certain) continuum
estimates.
}
\label{ew_hist}
\end{figure*}

\subsection{The Equivalent Width Distribution}
\label{ew}

The rest frame equivalent widths reported for each source in
Table~\ref{table_spec_prop} were determined directly from the spectra
according to $\wrst = (F_{\ell} / f_{\lambda,r}) / (1+z)$, where
$F_{\ell}$ is the flux in the emission line and $f_{\lambda,r}$ is the
measured red--side continuum flux density.  For the one case with
$f_{\lambda,r}$ formally consistent with zero, we derive a $2 \sigma$
lower limit to the equivalent width. The resulting equivalent width distribution is plotted in
Figure~\ref{ew_hist}, with individual measurements keyed to the grating
with which each source was observed. There is no obvious systematic
difference between equivalent widths measured in our lower resolution
spectroscopic configuration and those measured in our higher resolution
configuration.

Using Starburst99 models \citep{leitherer99} with a Salpeter initial
mass function (IMF), an upper
mass cutoff of 120 $M_{\sun}$, and a metallicity of 1/20th solar,
\citet{malhotra02} calculated maximum equivalent widths of 300 \AA\ for a
one Myr--old stellar population, 150 \AA\ for a 10 Myr--old population,
and 100 \AA\ for populations older than $10^8$ years. Owing to the lower
metallicity, these values are slightly higher than the previous
calculations of \citet{charlot93}, who present 240 \AA\ as the highest
equivalent width achievable by a stellar population.

To properly compare these model predictions to the equivalent width
distribution presented herein, one must first account for absorption in
the intergalactic medium (IGM), which affects the observations but is not
included in the models. Equivalent widths determined from spectroscopy are
based on the flux in the observed emission line (which may be affected by
intergalactic absorption), and on the red--side continuum flux density
(which is unaffected by intergalactic absorption).  \citet{malhotra02}
used the prescription for IGM absorption given by \citet{madau95} to
derive a flux decrement of a factor of 0.64 for an intrinsically symmetric
\lya\ emission line centered on zero velocity. Of course,
this calculation neglects the effects of the dust content
and detailed kinematics of the galaxy ISM, which increasingly
appear to play a significant role in determining the emergent \lya\
profile
\citep[e.g.][]{kunth98, stern99, mas-hesse03, shapley03, ahn04}.
Nonetheless, adopting this
correction factor as an upper limit on possible IGM effects dictates that
the limiting model equivalent widths measured spectroscopically would be
190 \AA, 100 \AA, and 60 \AA\ for populations of ages $10^6, 10^7$ and
$10^8$ years respectively.

A second concern regarding the equivalent width determination in
spectroscopy is the high sensitivity to uncertainty in the continuum
measurement.  Since the continuum estimate enters into the
denominator of the expression for \wrst, the characteristically small
continuum values, along with their considerable error bars, tend to cause
large scatter in the results.  Not surprisingly, the sources in our sample
with the largest measured equivalent widths tend also to have the largest
fractional uncertainty therein.

With these caveats in mind, we performed a careful statistical analysis
designed to place rigorous constraints
on the number of galaxies in our sample with equivalent widths in excess
of the maximum values allowed by stellar population models. First, we
associated each measured line flux $F_{\ell,i} \pm \delta F_{\ell,i}$ with
a Gaussian probability density function (PDF) centered on $F_{\ell,i}$
with width $\sigma = \delta F_{\ell,i}$; we proceeded similarly for the
measured continuum fluxes. We then generated a grid of line flux versus
continuum flux on which each node has an associated equivalent width and
is assigned a weight according to the Gaussian error distribution on
each of its fluxes.  Next we collapsed the grid into a histogram of
equivalent widths, adding the weight from each grid point to the
appropriate equivalent width bin.  The result is a non--Gaussian PDF $P_i
(w)$ for which $P_i (w) \, dw$ is the probability of observing \wrsti\ in
the interval $w < \wrsti\ < w + dw$.  We used the final ensemble of $P_i
(w)$ to determine the likelihood that exactly $N$ galaxies in our sample
exceed some limiting $\wrst$, and we added these likelihoods to determine
the confidence with which a range of $N$ galaxies exceed that $\wrst$
limit.

Proceeding in this fashion, we find with 90\% confidence that 3 to 5 galaxies in our sample
exceed the fiducial \citet{charlot93} upper limit of $\wrst > 240$ \AA,
and with 91\% confidence that 4 to 6 galaxies exceed $\wrst > 190$ \AA,
the upper limit of \citet{malhotra02}, with the largest reasonable
correction for IGM absorption. Thus, no
matter how one treats the effects of IGM absorption on the maximum
equivalent widths predicted by stellar population models, we find a
significant fraction of our sample in excess of those limits. These
galaxies are required to be very young (age $< 10^6$ years), or to have
skewed IMFs, or perhaps to harbor AGN producing stronger--than--expected
\lya\ emission.

\subsection{AGN Among the $z=4.5$ Population?}
\label{agn}

Given the large equivalent widths measured for the high--redshift \lya\
emission in this sample and elsewhere 
\citep{kudritzki00, rhoads03, rhoads04}, one intriguing scenario is the possibility
that our narrow band selection has identified a large population of
high--redshift AGN.  However, the recent non--detections in deep ($\sim
170$ ks) {\it Chandra}/ACIS imaging of the narrow band--selected sources in
the LALA \bootes\ field \citep{malhotra03} and the LALA Cetus field 
\citep{wang04} have placed strong constraints against AGN
activity among the \lya--emitters.  No individual \lya\ candidate was
detected with {\it Chandra}/ACIS to a 3$\sigma$ limiting X--ray luminosity
of $L_{2-8 \mbox{\tiny keV}} = 3.3 \times 10^{43}$ erg s$^{-1}$. The
constraint is even stronger in the stacked X--ray image, which suggests a
3$\sigma$ limit to the average X--ray luminosity of $L_{2-8 \mbox{\tiny
keV}} < 2.8 \times 10^{42}$ erg s$^{-1}$.  This limit is roughly an order
of magnitude fainter than what is typically observed for even the heavily
obscured, Type II AGN \citep[e.g.][]{stern02q,norman02,dawson03}.

The case against AGN activity among the \lya--emitters is further borne
out by the optical spectroscopy presented herein.  The narrow physical
widths of the \lya\ emission lines ($\Delta v < 500$ km s$^{-1}$)
definitively rule out conventional broad--lined (Type I) AGN, and also
disfavor narrow--lined (Type II) AGN, which have
typical $\Delta v_{\tiny \lya} \sim 1000$ km s$^{-1}$.  Furthermore, no
individual spectrum shows evidence of the high--ionization state UV
emission lines symptomatic of AGN activity (e.g.\ \nvw, \civw, \heiiw)
nor is there evidence of such lines in the composite spectra
(Figures~\ref{compy400} and \ref{compy150}).  In particular, \civw\ flux
in the $400\ell$/mm composite spectrum is constrained to be $\simlt 8\%$ ($2 \sigma$) of
the flux in \lya, implying a flux ratio of $f(\lya)/f(\civw) \simgtr 13$.
By contrast, the three Type II AGNs cited above span only $0.7 <
f(\lya)/f(\civw) < 5.4$.

\subsection{Population III Among the $z=4.5$ Population?}
\label{heii}

The identification of a population of large equivalent width
\lya--emitters evidently powered by star formation in low
metallicity gas suggests that we are closing the gap between the first,
little--enriched primordial galaxies and the higher metallicities of
massive galaxies in the local universe. Indeed, recent numerical studies
of the rest frame UV and optical properties of very low metallicity
stellar populations indicate that \lya\ emission increases strongly with
decreasing metallicity, far exceeding $\wrst \simgtr 500$ \AA\ for $Z <
10^{-5} Z_{\sun}$ \citep[e.g.][]{schaerer03}. The tantalizing limit of
such studies is star formation in zero metallicity gas, the so--called
Population III, which constitutes the first bout of star formation in the
pre--galactic Universe.

The striking features of massive Population III stars are their high
effective temperatures ($T_{\mbox{\tiny eff}} \sim 10^5$ K for $M > 100
M_{\sun}$) and consequent hard ionizing spectra, resulting in the
production of 60\% more \ion{H}{1}--ionizing photons than their Population II
counterparts, and up to $10^5$ times more \ion{He}{2}--ionizing photons
\citep{tumlinson03}. As a consequence, in addition to high equivalent
width \lya\ emission, a unique observational signature of this primeval
population is emission from \heii\ recombination lines.  These lines are particularly
attractive for a detection experiment since they suffer minimal effects
of scattering by gas and benefit from decreasing attenuation by dust.
That said, the possibility for the direct detection of \heiiw\ in the
present data set boils down to two questions:  Is Population III star--formation
occurring at the comparatively late epoch occupied by the $z \approx 4.5$
\lya--emitters?  And if so, is the resulting \ion{He}{2} emission of sufficient
luminosity to be detected?

As for the first question, if star formation feedback in the Universe
is sufficiently weak that metal production proceeds very inhomogeneously,
then Population III star formation may continue to surprisingly 
low redshifts, occurring in regions that have not yet been polluted
by previous episodes of star formation.  To this end, the analytical models
of \citet{scannapieco03} designed to investigate the detectability of primordial star formation
indicate that Population III objects tend to form in the $10^{6.5}$--$10^7
M_{\sun}$ mass range, just large enough to cool within a Hubble time, but
small enough that they are not clustered near areas of previous star
formation.  The result is that somewhere between
1\% and 30\% of strong \lya--emitters at $z=4.5$ should owe their
\lya\ emission to Population III star formation, depending on the 
strength of feedback in these systems.

As for the detectability of the resulting \ion{He}{2} emission,
early predictions of \lya\ and \heii\ recombination
luminosities in metal--free stellar clusters suggest \heiiw\ emission in
excess of 13\% that of \lya, with $\wrst (\heiiw) \sim 1100$ \AA\
\citep{bromm01}.  Notably, \citet{schaerer02} points out that the
prediction of such a large equivalent width likely results from neglecting
the contribution of nebular continuum emission to the model Population III
spectrum, which dominates the SED for $\lambda > 1400$ \AA\ and therefore
reduces the expected equivalent width.
More recent models are indeed more conservative in their predictions of
the strength of \heii\ emission. \citet{tumlinson03} suggest \heii\ fluxes
for a Population III cluster at $z \approx 4.5$ ranging from $\sim$
0.001\% to 5\% of the flux in \lya, where the low extremum is set by the
limiting case of an instantaneous starburst, and the high extremum is set by
constant star formation at rate of $\sim 40$ $M_{\sun}$ yr$^{-1}$. 
\citet{schaerer03} predicts \heiiw\ equivalent widths in
excess of 80 \AA\ for very young zero--metallicity instantaneous
starbursts, though at ages greater than $\sim 1$ Myr, values $\wrst
(\heiiw) \simgtr 5$ \AA\ are expected only at metallicities $Z < 10^{-7}
Z_{\sun}$.

Physically, the best prospect for detecting \heiiw\ in our
spectroscopic sample lies with the spectra containing \lya\ at the highest
equivalent widths. However, \heiiw\ at the suggested flux levels is
undetectable in any of our individual spectra, leaving only the
possibility that a constraint to \heiiw\ emission may be derived from the
composite spectrum, where we benefit from a $\sqrt{N}$--reduction in
Poission noise. Accordingly, we performed Monte Carlo simulations aimed at
measuring \heiiw\ emission in the $400\ell$/mm composite, searching over a
distribution of potential \heiiw\ line widths set by the width of the
composite \lya\ line.  The result is a \heiiw\ flux which is formally
consistent with zero, with a 2$\sigma$ (3$\sigma$) upper limit of 13\%
(20\%) of the flux in \lya. The corresponding 2$\sigma$ (3$\sigma$) upper
limit to the \heiiw\ rest frame equivalent width is 17 \AA\ (25 \AA). This
equivalent width limit is sufficient to rule out the youngest
zero--metallicity instantaneous burst and continuous SFR models of
\citet{schaerer03}, though metallicities of $Z < 10^{-7} Z_{\sun}$ are
still permissible. We therefore conclude that this data set
cannot corroborate the proposition \citep[e.g.][]{tumlinson03} that the high
equivalent widths of the $z \approx 4.5$ narrow band--selected
\lya--emitters suggest that the first metal--free stars have already been
found.

\section{Conclusion}
\label{conclusion}

Out of 25 narrow band--selected candidates, we have spectroscopically
confirmed 18 galaxies at $z \approx 4.5$, implying a selection
reliability of 72\%. The resulting sample of confirmed \lya\ emission
lines show large equivalent widths (median $\wrst \approx 80$ \AA) but
narrow physical widths ($\Delta v < 500$ km s$^{-1}$), supporting the
conclusion of \citet{malhotra03} and \citet{wang04} that
the \lya\ emission in these sources derives from star formation, not from
AGN activity.  Moreover, though the expectation from theoretical models of
galaxy formation in the primordial Universe is that a small fraction of
\lya--emitting galaxies at $z \approx 4.5$ may be nascent, metal--free
objects, we do not detect \heiiw\ emission in either individual or
composite spectra, indicating that though these galaxies are young, they
show no evidence of being truly primitive, Population III objects. Of
course, this last result may be a function of our comparatively small
sample size, which could only be reasonably expected to yield a \heiiw\
detection if the frequency of Population III objects among $z \approx 4.5$
\lya--emitters exceeds $\sim 6\%$.  Clearly, increasing our sample size
with future spectroscopy will provide a
far tighter constraint on the make--up of the $z \approx 4.5$ galaxy
population.

One heretofore unexplored consideration is the possibility that
galactic--scale winds may be required to work in concert with sub--solar
metallicities to facilitate the escape of \lya\ radiation from a system
\citep[e.g.][]{kunth98, ahn04}. In this scenario, the low metallicity acts to
reduce the dust opacity, and the wind acts to Doppler shift the absorbers,
minimizing resonant scattering of \lya\ photons.  The detailed geometry of
the interstellar medium (ISM) doubtless plays several roles in this
process.  As one example, a galactic wind driven by star formation has its
origin in an over--pressured cavity of hot gas inside the star--forming
galaxy.  This superbubble ultimately expands and bursts out into the
galaxy halo; naturally this expansion, and hence the galactic wind,
proceeds in the direction of the vertical pressure gradient
\citep{heckman00,tenorio99}. As a second example, if dust and neutral gas
in the ISM have a high covering factor but a low volume filling factor, it
is possible for continuum radiation to be strongly absorbed while an
appreciable fraction of \lya\ line radiation escapes \citep{neufeld91}.

Observationally, the extent to which the emission of \lya\ photons in a   
star forming galaxy depends on not only the distribution and kinematics of
gas and dust in its ISM, but also on the inclination of the system,
remains an open question. Since significant \lya\ emission in SCUBA
sources \citep{chapman03} offers evidence that the deleterious effect of
dust on \lya\ emission may be mitigated by strong starburst--driven winds, 
it appears unlikely that spectroscopy of the rest frame UV of high
redshift \lya--emitters alone will be able to fully disentangle these
effects. By contrast, a detailed understanding of the rest frame optical
properties of the these systems would offer a strong lever arm on their
dust contents and star formation histories.  Hence, resolution of these
issues may await infrared observations of the $z \approx 4.5$ galaxy
sample.

%%%%%%%%%%%%%%%%%%%
% Acknowledgments %
%%%%%%%%%%%%%%%%%%%

\acknowledgements

This work benefited greatly from conversations with T. Robishaw, J. Simon,
D. Schaerer, E. Green, and the anonymous referee.  
We further acknowledge J.~G. Cohen and C.~C. Steidel for
supporting LRIS--R and LRIS--B, respectively, and we thank the DEEP2 team
for providing the sample \oiiw\ spectra. In addition, we wish to
acknowledge the significant cultural role that the summit of Mauna Kea
plays within the indigenous Hawaiian community; we are fortunate to have
the opportunity to conduct observations from this mountain. The work of SD
was supported by IGPP--LLNL University Collaborative Research Program
grant \#03--AP--015, and was performed under the auspices of the U.S.
Department of Energy, National Nuclear Security Administration by the
University of California, Lawrence Livermore National Laboratory under
contract No.\ W--7405--Eng--48.  The work of DS was carried out at the Jet
Propulsion Laboratory, California Institute of Technology, under contract
with NASA.  AD and BJ acknowledge support from NOAO, which is operated by the 
Association of Universities for Research in Astronomy, Inc. under 
cooperative agreement with the National Science Foundation (NSF).
HS gratefully acknowledges NSF grant AST 95--28536 for
supporting much of the research presented herein. This work made use of
NASA's Astrophysics Data System Abstract Service.

%%%%%%%%%%%%%%
% References %
%%%%%%%%%%%%%%

\eject

\end{document}